\documentclass[twocolumn]{aastex63}

\usepackage[version=4]{mhchem}
\usepackage{multirow}
\received{}
\revised{}
\accepted{}
\submitjournal{ApJ}

\shorttitle{HC$^{18}$O$^{+}$ in TW Hya}
\shortauthors{Furuya et al.}

\newcommand{\ab}[1]{x({\rm #1})}

\begin{document}

\title{Detection of HC$^{18}$O$^{+}$ in a protoplanetary disk: exploring oxygen isotope fractionation of CO}

\correspondingauthor{Kenji Furuya}
\email{kenji.furuya@nao.ac.jp}
\correspondingauthor{Takashi Tsukagoshi}
\email{Takashi.Tsukagoshi@nao.ac.jp}

\author{Kenji Furuya}
\affiliation{National Astronomical Observatory of Japan, Osawa 2-21-1, Mitaka, Tokyo 181-8588, Japan}
\affiliation{These authors contributed equally to this work.}
\author{Takashi Tsukagoshi}
\affiliation{National Astronomical Observatory of Japan, Osawa 2-21-1, Mitaka, Tokyo 181-8588, Japan}
\affiliation{These authors contributed equally to this work.}
\author{Chunhua Qi}
\affiliation{Harvard-Smithsonian Center for Astrophysics 60 Garden Street Cambridge, MA 02138, USA}
\author{Hideko Nomura}
\affiliation{National Astronomical Observatory of Japan, Osawa 2-21-1, Mitaka, Tokyo 181-8588, Japan}
\author{L. Ilsedore Cleeves}
\affiliation{Department of Astronomy, University of Virginia, Charlottesville, VA 22904, USA}
\author{Seokho Lee}
\affiliation{National Astronomical Observatory of Japan, Osawa 2-21-1, Mitaka, Tokyo 181-8588, Japan}
\author{Tomohiro C. Yoshida}
\affiliation{Department of Astronomical Science, The Graduate University for Advanced Studies (SOKENDAI),
Osawa, Mitaka, Tokyo 181-8588, Japan}
\affiliation{National Astronomical Observatory of Japan, Osawa 2-21-1, Mitaka, Tokyo 181-8588, Japan}



\begin{abstract}
The oxygen isotope fractionation scenario, which has been developed to explain the oxygen isotope anomaly in the solar system materials,
predicts that CO gas is depleted in \ce{^18O} in protoplanetary disks,
where segregation between solids and gas inside disks had already occurred.
Based on ALMA observations, we report the first detection of \ce{HC^18O+}(4--3)
in a Class II protoplanetary disk (TW Hya).
This detection allows us to explore the oxygen isotope fractionation of CO in the TW Hya disk
from optically thin \ce{HCO+} isotopologues as a proxy of optically thicker CO isotopologues.
Using the \ce{H^13CO+}(4--3) data previously obtained with SMA, we find that the \ce{H^13CO+}/\ce{HC^18O+} ratio in the central $\lesssim$100 au regions of the disk is $10.3\pm3.2$.
We construct a chemical model of the TW Hya disk with carbon and oxygen isotope fractionation chemistry, 
and estimate the conversion factor from \ce{H^13CO+}/\ce{HC^18O+} to \ce{^13CO}/\ce{C^18O}.
With the conversion factor (= 0.8), the \ce{^13CO}/\ce{C^18O} ratio is estimated to be $8.3 \pm 2.6$, which is consistent with the elemental abundance ratio in the local ISM ($8.1\pm0.8$) within error margin.
Then there is no clear evidence of \ce{^18O} depletion in CO gas of the disk, although we could not draw any robust conclusion due to uncertainties.
In conclusion, optically thin lines of \ce{HCO+} isotopologues are useful tracers of CO isotopic ratios, which are hardly constrained directly from optically thick lines of CO isotopologues. 
Future higher sensitivity observations of \ce{H^13CO+} and \ce{HC^18O+} would be able to allow us to better constrain the oxygen fractionation in the disk.
\end{abstract}





\section{Introduction}
\label{sec:intro}
Molecular isotope ratios, such as D/H ratio of water, are powerful tools to understand the origin of solar system
materials and their possible chemical link with interstellar materials \citep[e.g.,][]{cleeves14,furuya17}.
The element oxygen has three stable isotopes: \ce{^16O}, \ce{^17O}, and \ce{^18O}.
It has been well established that the solar system materials, including chondrules, Ca-Al-rich inclusions (CAIs), 
and the Earth's ocean, are enriched in \ce{^17O} and \ce{^18O} compared to the Sun, 
and their oxygen isotope compositions show mass-independent variations \citep[e.g.,][]{tenner18}.
Fe-O-S-bearing cosmic symplectites within Acfer 094 meteorite shows the most significant $^{17,18}$O enrichment among the solar system materials;
the \ce{^16O}/\ce{^17O} and \ce{^16O}/\ce{^18O} ratios are larger 
by 25 \% compared to the Sun \citep{sakamoto07}.

The leading hypothesis to explain the oxygen isotope fractionation observed in the solar system materials 
is isotope-selective photodissociation of CO by ultraviolet (UV) photons, 
either in the parent molecular cloud or in the protosolar disk, followed by the formation of  $^{17,\,18}$O-enriched water ice \citep[e.g.,][]{clayton02,yurimoto04,lyons05}.
Recent detailed study of CAIs suggests that the parent molecular cloud is the more favorable place for that chemistry \citep{krot20}.
CO photodissociation is subject to self-shielding (i.e., CO itself can be the dominant absorber of dissociating UV radiation rather than dust grains).
Because rarer isotopologues (e.g., \ce{C^18O} and \ce{C^17O}) are less abundant than \ce{^12CO}, 
they are not self-shielded until deeper into the cloud or the disk.
This property makes CO photodissociation an isotope-selective process \citep[e.g.,][]{visser09,miotello14}.
$^{17,\,18}$O-enriched atomic oxygen is produced as the photodissociation product of CO, and can be converted into $^{17,\,18}$O-enriched water ice.
As a result, the three major oxygen reservoirs in star- and planet-forming regions, CO gas, \ce{H2O} ice, and refractory silicate, could have different isotope compositions, where, for example, 
(i) CO gas is depleted in $^{17,\,18}$O, (ii) \ce{H2O} ice is enriched in $^{17,\,18}$O, and (iii) silicate dust shows no fractionation.
In our solar system, \ce{H2O} in cometary ice is enriched in \ce{^18O}, which is consistent with this scenario \citep[e.g.,][]{schroeder19,altwegg19}.
In the protosolar disk, dust grains coated with water ice could settle to the midplane and drift radially inward 
(i.e., segregating $^{17, 18}$O-rich solids and $^{17, 18}$O-poor gas inside the disk).
When the icy dust grains across the snowline, water ice sublimates \citep[e.g.,][]{cuzzi04}. 
The water vapor could react with silicates to make solid materials in the inner solar system enriched in \ce{^17O} and \ce{^18O} \citep[e.g.,][]{yurimoto04,yamamoto18}, 
which could be incorporated into planetasimals (asteroids/meteorites) and ultimately into planets.

The oxygen isotope fractionation scenarios outlined above has been favored to explain the oxygen isotope compositions in the solar system materials,
but it remains unclear whether this evolution can occur in all star-forming regions or some specific factors are required.
For example, based on the physico-chemical model of the collapsing proto-solar cloud, \citet{lee08} suggested that 
UV radiation field enhanced by a nearby massive star is necessary to explain the oxygen isotope compositions in the solar system materials.
To address the question, studies of oxygen isotope fractionation in both star- and planet-forming regions are crucial.
It is, however, not straightforward to constrain the oxygen isotope fractionation in Class II disks, 
because of the faint (or the non-detection of) H$_2$O emission \citep{hogerheijde11,du17,vandishoeck21},
and because \ce{^12C^16O} and \ce{^13C^16O} rotational lines are often optically thick \citep{schwarz16,huang18}.
To the best of our knowledge, \citet{smith09} is the only study, which observationally constrains the oxygen isotope fractionation of CO in protoplanetary disks.
They reported that  the \ce{^12C^16O}/\ce{^12C^18O} ratio in a disk in VVCra, a binary T Tauri star, is $690 \pm 30$ based on the observations of 
rovibrational absorption lines of the CO isotopologues.
The derived \ce{^12C^16O}/\ce{^12C^18O} ratio is higher than the elemental abundance ratio [\ce{^16O}]/[\ce{^18O}] in the local ISM of $557\pm30$ \citep{wilson99},
probably due to the isotope selective photodissociation of CO in the disk surface \citep{smith09}.



Here we report the first detection of \ce{HC^18O+} in a Class II protoplanetary disk around a T Tauri star, TW Hya.
The TW Hya disk is one of the most studied Class II disks in terms of both physical and chemical structures.
With the previously detected optically thin \ce{H^13CO+} emission \citep{qi08}, 
this detection allows us to constrain the oxygen isotope fractionation of CO in the TW Hya disk,
using a \ce{H^13CO+}/\ce{HC^18O+} ratio as a proxy of a \ce{^13CO}/\ce{C^18O} ratio.
According to previous studies \citep{schwarz16,huang18}, 
\ce{CO}(3-2) and \ce{^13CO}(3-2) transitions are optically thick in the entire disk.
Even \ce{^18CO}(3-2) is optically thick in the inner disk \citep[inside $\sim$20 au;][]{zhang17}.
Thus, it is hard to constrain the oxygen isotope ratio of CO in the TW Hya disk directly 
from the observations of CO isotopologues.

The TW Hya disk is an ideal lab to search for isotopic variations in C and O, 
as it is known to have a low CO and \ce{H2O} abundances, 
based on observations of these species and HD \citep[e.g.,][]{hogerheijde11,bergin13,zhang19}.
The underabundance of volatile carbon and oxygen could be a consequence of dust evolution and segregation between solids and gas inside the disk \citep{du15,krijt18}.
Then, the segregation of solids and gas are likely common process to cause the volatile evolution observed in the TW Hya disk
and the oxygen isotope fractionation observed in the solar system materials.
This makes the TW Hya disk an ideal place to test the oxygen isotope fractionation theory;
if the oxygen fractionation mechanism is/was at work in the TW Hya disk as in the proto-solar disk, CO in TW Hya should be depleted in \ce{^18O} (and \ce{^17O}).

The rest of this paper is organized as follows.
Section \ref{sec:obs} summarizes the observational details and the data reduction procedure.
Section \ref{sec:result} presents the observational data products and the disk-integrated \ce{H^13CO+}/\ce{HC^18O+} ratio.
In Section \ref{sec:model}, we discuss the relation between the \ce{^13CO}/\ce{C^18O} ratio and the \ce{H^13CO+}/\ce{HC^18O+} ratio 
with the help of thermo-chemical model of the TW Hya disk.
The interpretation of our findings is discussed in Section \ref{sec:discuss}.


\section{Observations} \label{sec:obs}
We analyzed the ALMA archive data (ID:2016.1.00311.S) to search for the \ce{HC^18O+} emission line from the protoplanetary disk around TW~Hya.
The TW Hya disk is almost face-on with the inclination of 7$^\circ$ and the distance from the earth is 59.5 pc \citep{gaia16}.
The data consisted of two execution blocks; observations with an array configuration of C40-1 on 2017 April 8 and C40-5 on May 21.
The integration times for the execution blocks were 28.8~min and 47.8~min for C40-1 and C40-5, respectively.
The Band~7 receiver system was employed to detect the continuum emission at 347~GHz and some molecular lines.
The 14 spectral windows (SPWs) were used in the Frequency Division Mode, and one of which was tuned for detecting the \ce{HC^18O+}(4--3) transition.
The bandwidth and the channel spacing were 58.5938~MHz and 61.035~kHz, respectively, corresponding to a velocity resolution of 54~m~s$^{-1}$.
The line-free channels of the SPWs were used for detecting the continuum emission at Band~7, whose aggregate bandwidth was $\sim$1.8~GHz.
A quasar J1037-2934 was used for the calibration of the complex gain of the visibilities as well as the calibration for the bandpass characteristics.

The visibility data were reduced and calibrated using the Common Astronomical Software Application package \citep{mcmullin07}.
The pipeline script given by ALMA was applied for each execution block for the initial data flagging and the calibration for the bandpass characteristics, complex gain, and flux scaling.
After the flagging of bad data manually, the molecular line and the continuum were separated using {\it uvcontsub}.
The continuum emission was taken from the line-free channel in all the SPWs employed in this observation.

Before the imaging of the \ce{HC^18O+} line, we first made a continuum map from the visibilites of two execution blocks for making a calibration table from the self-calibration.
We created the dirty map for each execution block, and to concatenate two execution blocks the disk center was determined by a 2D Gaussian fitting to the bright part of the disk emission using {\it imfit}.
After that, the center of the field of view was modified to the disk center by {\it fixvis}, and we concatenated two execution blocks using {\it concat} with a positional tolerance.
Then, the CLEAN map of the dust emission was reconstructed from the concatenated visibilities using {\it tclean}.
For the imaging, we used the Briggs weighting with a robust parameter of 0.5.
We also employed the multiscale CLEAN algorithm with scale parameters of [0, 0.2, 0.6] arcsec.
With the initial CLEAN map as a model, we applied the self-calibration for the continuum data.
We first made a calibration table solved in phase for the whole part of each execution block, and then the solution intervals were shortened from 300, 60, and 6~seconds for the short baseline execution block and from 300, 180, and 120~seconds for the long baseline one with iteratively improving the model visibility by making the CLEAN map. 
After the phase-only self-calibration described above, we applied the self-calibration in amplitude with a solution interval of the whole part of each execution block.
Finally, the beam size of the continuum CLEAN map was $0\farcs23\times0\farcs20$ with a position angle of $80\fdg3$, and the sensitivity was 95~$\mu$Jy~beam$^{-1}$.

The \ce{HC^18O+} emission line was imaged from the continuum-subtracted measurement set after applying the calibration table obtained by the self-calibration for the continuum emission.
The CLEAN map was reconstructed using {\it tclean} with the Briggs weighting (robust$=$0.5).
To improve the image sensitivity, we employed a uvtaper of 0.8~arcsec.
The multiscale CLEAN was used with scale parameters of [0, 0.85, 2.55]~asec.
The velocity width of the image cube was gridded to 0.06~km~s$^{-1}$, which is almost comparable to the velocity resolution of the channels.
The beam size of the CLEANed \ce{HC^18O+} map was $1\farcs05\times0\farcs86$ with a position angle of $85\fdg1$.
The noise level of the channel map was 75.2~mJy~beam$^{-1}$.
The observed line properties are summarized in Table \ref{table:line}.

%

\begin{table*}
\caption{Summary of observed lines}
\begin{center}
\begin{tabular}{ccccccc}
\hline\hline        
Transition & Rest Frequency & $E_u$ & Beam (PA) & Flux $< 4.1''\times1.8''$ \tablenotemark{{\rm b}}  & Flux $< 5''\times5''$  \tablenotemark{{\rm c}} & Reference \\
               & (GHz)              & (K)    &              &  (Jy km s$^{-1}$)  &   (Jy km s$^{-1}$)                  &\\  
\hline
\ce{HC^18O+}(4--3) & 340.6329780 & 40.87 & $1\farcs05\times0\farcs86$ ($85\fdg1$)  &  $0.07 \pm 0.01$ &  $0.14 \pm 0.01$ &This work \\
\ce{H^13CO+}(4--3) & 346.9983381 & 41.63 & $4\farcs1\times1\farcs8$  ($3\fdg3$\tablenotemark{{\rm a}}) & $0.76 \pm 0.16$ & $1.8 \pm 0.2$ &\citet{qi08}  \\
\hline
\end{tabular}
\end{center}
\tablecomments{
$^{a}$Value for the central $4.1''\times1.8''$ region.
$^{b}$ Uncertainty does not include flux calibration uncertainty. The integrated flux of \ce{H^13CO+}(4--3) is slightly different from that presented in 
\citet[][0.61 Jy km s$^{-1}$]{qi08}, because we recalculated it using the same procedure as the calculation of the \ce{HC^18O+}(4--3) integrated flux.
$^{c}$ Uncertainty does not include flux calibration uncertainty.
}
\label{table:line}
\end{table*}

\section{Results} \label{sec:result}
\subsection{HC$^{18}$O$^+$ emission map and spectra}
\begin{figure}[t]
    \plotone{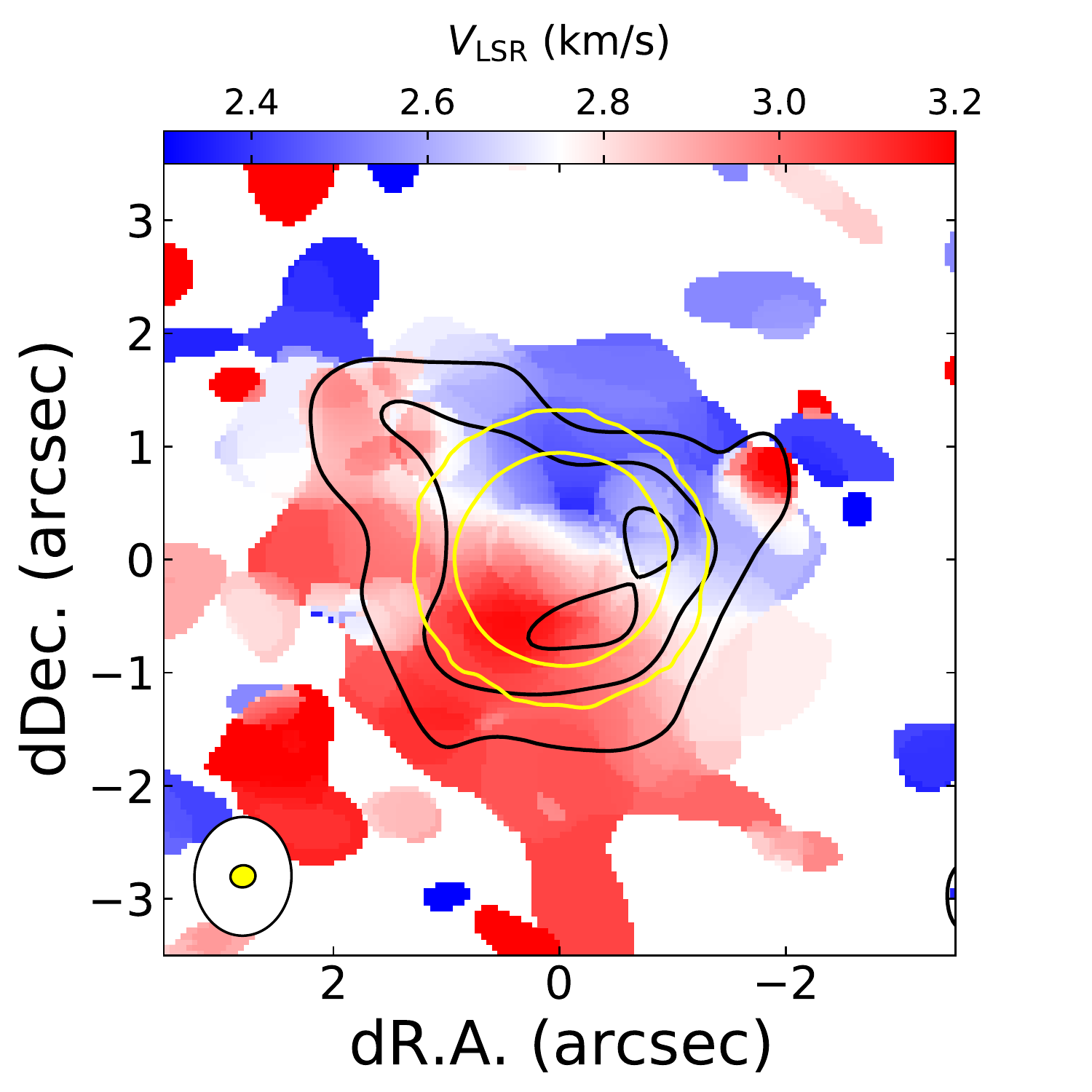}
    \caption{Map of the \ce{HC^18O+}(4--3) emission from the protoplanetary disk around TW~Hya. The color scale indicates the weighted-mean velocity map. The integrated intensity from 2.3 to 3.5~km~s$^{-1}$ is shown by the black contour. The contour interval is 2$\sigma$ starting from 2$\sigma$, where $1\sigma=3.1$~mJy~km~s$^{-1}$. The yellow contour represents the 10 and 100$\sigma$ boundaries of the continuum emission, where $1\sigma=92$~$\mu$Jy~km~s$^{-1}$. The ellipse at the bottom-left corner denotes the beamsize of the map, in which the \ce{HC^18O+} is shown in white and the continuum is in yellow, respectively.}
    \label{fig:map}
\end{figure}

The \ce{HC^18O+}(4-3) emission is associated with the protoplanetary disk of TW~Hya as shown in Figure\ \ref{fig:map}.
The position angle of the velocity gradient is consistent with previous CO measurements \citep{teague2019}, indicating that the \ce{HC^18O+} emission traces the Kepler rotating gas disk.
The emission peak is offset from the continuum peak and is slightly stronger on the west side;
these features are also seen in the \ce{HCO+}(3--2) emission in this disk \citep{cleeves15}.
The peak intensity of  \ce{HC^18O+} is $>$6 $\sigma$, where 1$\sigma$ = 3.1 mJy km s$^{-1}$ is the rms measured in the integrated intensity map.

\begin{figure}[ht!]
    \plotone{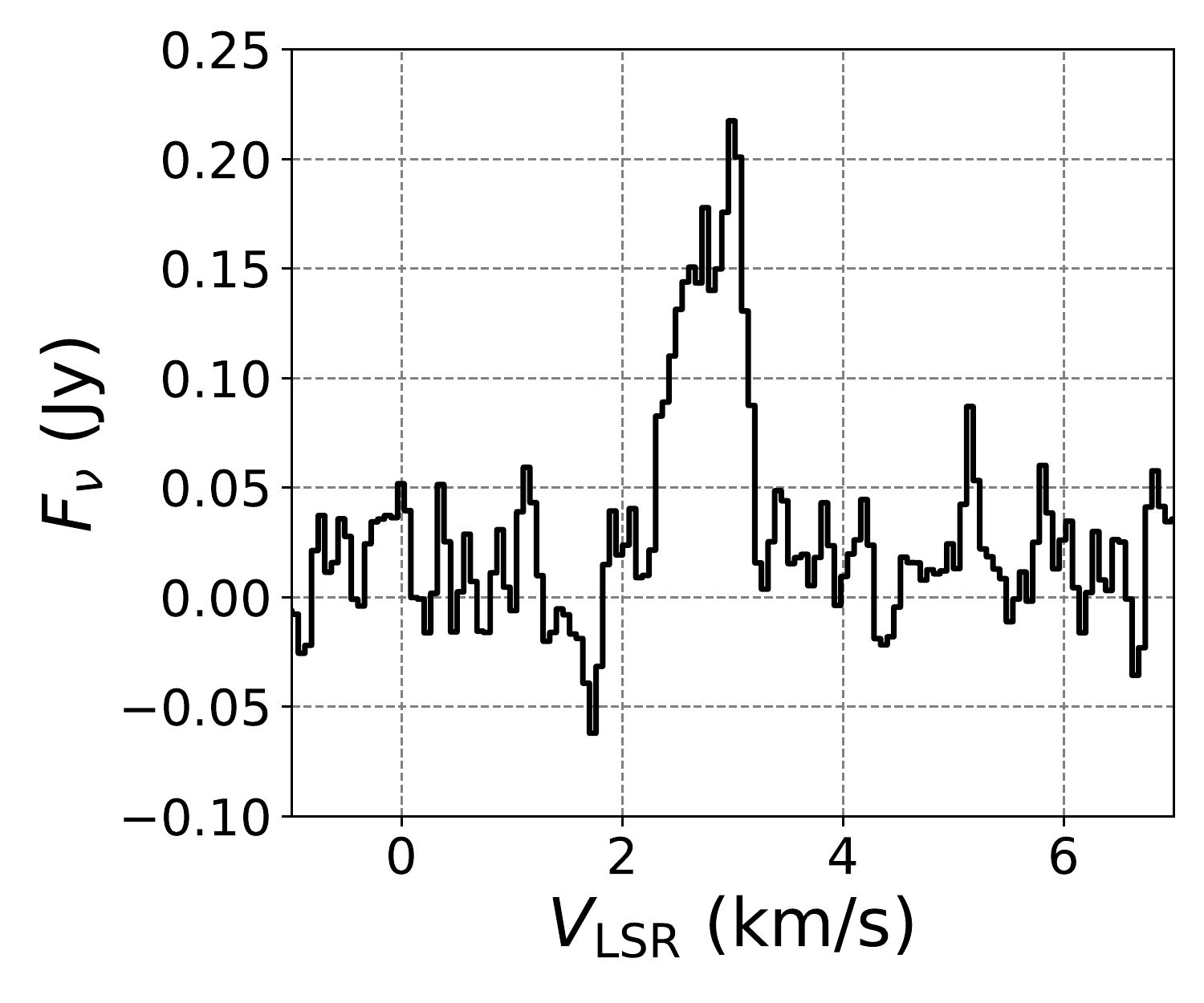}
    \caption{Spectrum of the flux density in \ce{HC^18O}(4--3) integrated over the central $5\arcsec\times5\arcsec$ box region.}
    \label{fig:spectrum}
\end{figure}

The spatially integrated spectrum of \ce{HC^18O+}(4--3) over the central $5''\times5''$  region is shown in Figure \ref{fig:spectrum}.
The spectrum shows the characteristic double-peak structure due to the Keplerian rotation.

\subsection{H$^{13}$CO$^+$/HC$^{18}$O$^+$ ratio}
Previous studies have found that \ce{HCO+}(4--3) emission in the disk is likely optically thick,
while \ce{H^13CO+}(4--3) emission is optically thin \citep{cleeves15}.
\ce{H^13CO+}(4--3) emission at 347 GHz was detected toward the TW Hya disk in the SMA observations 
with a beam of $4.1''\times1.8''$ \citep{qi08}.
We checked the ALMA archive and confirmed that there is no available data for \ce{H^13CO+}(4-3) and the other transitions of \ce{H^13CO+}.
Here we derive the H$^{13}$CO$^+$/HC$^{18}$O$^+$ ratio in the disk using the \ce{H^13CO+}(4-3) data 
obtained with SMA
and its interpretation will be discussed in the following sections.

Table \ref{table:line} lists the total integrated intensities of \ce{HC^18O+}(4--3) and \ce{H^13CO+}(4-3) 
within the central $4.1''\times1.8''$ region and across the entire \ce{HC^18O+} disk (within a radius of 2.5'').
The total integrated intensities in the central $4.1''\times1.8''$ region are converted into column densities using Eq. 82 in \citet{mangum15},
assuming LTE and optically thin conditions.
The parameters of the observed transitions were taken from the Cologne Database for Molecular Spectroscopy \citep{muller01,muller05}.
Assuming that excitation temperature ($T_{ex}$) is 30 K, which is a typical temperature of warm molecular layers \citep[e.g.,][see also Section \ref{sec:model}]{aikawa02},
the column densities of \ce{H^13CO+} and \ce{HC^18O+} in the central $4.1''\times1.8''$ region 
are ($5.8\pm1.5) \times10^{11}$ cm$^{-2}$ and $(5.6\pm1.0) \times10^{10}$ cm$^{-2}$, respectively.
The uncertainty includes 15 \% flux calibration uncertainty for the SMA data \citep{qi08} and 
10 \% flux calibration uncertainty for the ALMA data.
The flux calibration uncertainties do not significantly affect the uncertainties in the derived column densities, because the RMS noises are larger than the calibration uncertainties.
If $T_{ex}$ is assumed to be 20 K, the estimated column densities become slightly higher:
($7.8\pm2.0) \times10^{11}$ cm$^{-2}$ for \ce{H^13CO+} and $(7.5\pm1.3) \times10^{10}$ cm$^{-2}$ for \ce{HC^18O+}.

Then the \ce{H^13CO+}/\ce{HC^18O+} column density ratio in the central $4.1''\times1.8''$ region is estimated to be $10.3\pm3.2$, 
which is consistent with the elemental abundance ratio [\ce{^13C}][\ce{^16O}]/([\ce{^12C}][\ce{^18O}]) 
in the local ISM of $8.1\pm0.8$ \citep{wilson99} within error margin.
Note that the column density ratio is not sensitive to the choice of the excitation temperature, 
because the difference in upper state energy is less than 1 K between \ce{H^13CO+}(4--3) and \ce{HC^18O+}(4--3) (Table \ref{table:line}).

We also calculate the \ce{H^13CO+}/\ce{HC^18O+} column density ratio in the entire \ce{HC^18O+} disk (within a radius of 2.5'') in a similar way.
The ratio is evaluated to be $12.5\pm2.8$, which is consistent with that in the central $4.1''\times1.8''$ region within error margin ($10.3\pm3.2$).
Then, in the rest of this paper, we discuss the \ce{H^13CO+}/\ce{HC^18O+} column density ratio in the central $4.1''\times1.8''$ (240  au $\times$ 110  au) region.

\section{Disk chemistry modeling} \label{sec:model}
In this section, we discuss the relation between the \ce{^13CO}/\ce{C^18O} ratio and the \ce{H^13CO+}/\ce{HC^18O+} ratio. 
As \ce{HCO+} is formed by the proton transfer reaction, CO + H$_3^+$ $\rightarrow$ \ce{HCO+} + \ce{H2},
and is destructed by the dissociative recombination reactions with electrons, 
\ce{HCO+} + e$^-$ $\rightarrow$ CO + H,
one may expect that the isotopic ratios of \ce{HCO+} directly reflect those of CO.
According to previous astrohemical models in molecular clouds and protoplanetary disks \citep[e.g.,][]{langer84,woods09,furuya11}, this is not always the case; 
the isotopic ratios of \ce{HCO+} are modified by isotope exchange reactions.
CO and \ce{HCO^+} exchange carbon and oxygen isotopes via the following reactions \citep{mladenovic14,mladenovic17,loison19}:
\begin{align}
&{\rm H^{12}CO^+} + {\rm ^{13}CO} \rightleftharpoons {\rm H^{13}CO^+} + {\rm ^{12}CO} + 17.4\,\,{\rm K}, \label{react:13c} \\
&{\rm HC^{16}O^+} + {\rm C^{18}O} \rightleftharpoons {\rm HC^{18}O^+} + {\rm C^{16}O} + 6.3\,\,{\rm K}. \label{react:18o}
\end{align}
The forward reactions are exothermic, while the reverse reactions are slightly endothermic 
due to the difference in the zero point energy among the different isotopologues \citep{mladenovic17}.
\citet{woods09} found that the \ce{HCO+}/\ce{H^13CO+} abundance ratio deviates from \ce{CO}/\ce{^13CO}, 
and is determined by the balance between the forward and backward directions of Reaction \ref{react:13c} in their disk chemical model.
Note that the gas-phase CO abundance is the canonical value ($\sim$10$^{-4}$) in the disk model by \citet{woods09},
while in the TW Hya disk, it is lower than the canonical value by a factor of $>$10, 
based on the observations of HD and CO isotopologues \citep[e.g.,][]{favre13}.
Then it remains unclear whether the isotope exchange reactions are efficient enough to determine the isotopic ratios of \ce{HCO+} in the TW Hya disk.

\begin{figure}[t]
    \plotone{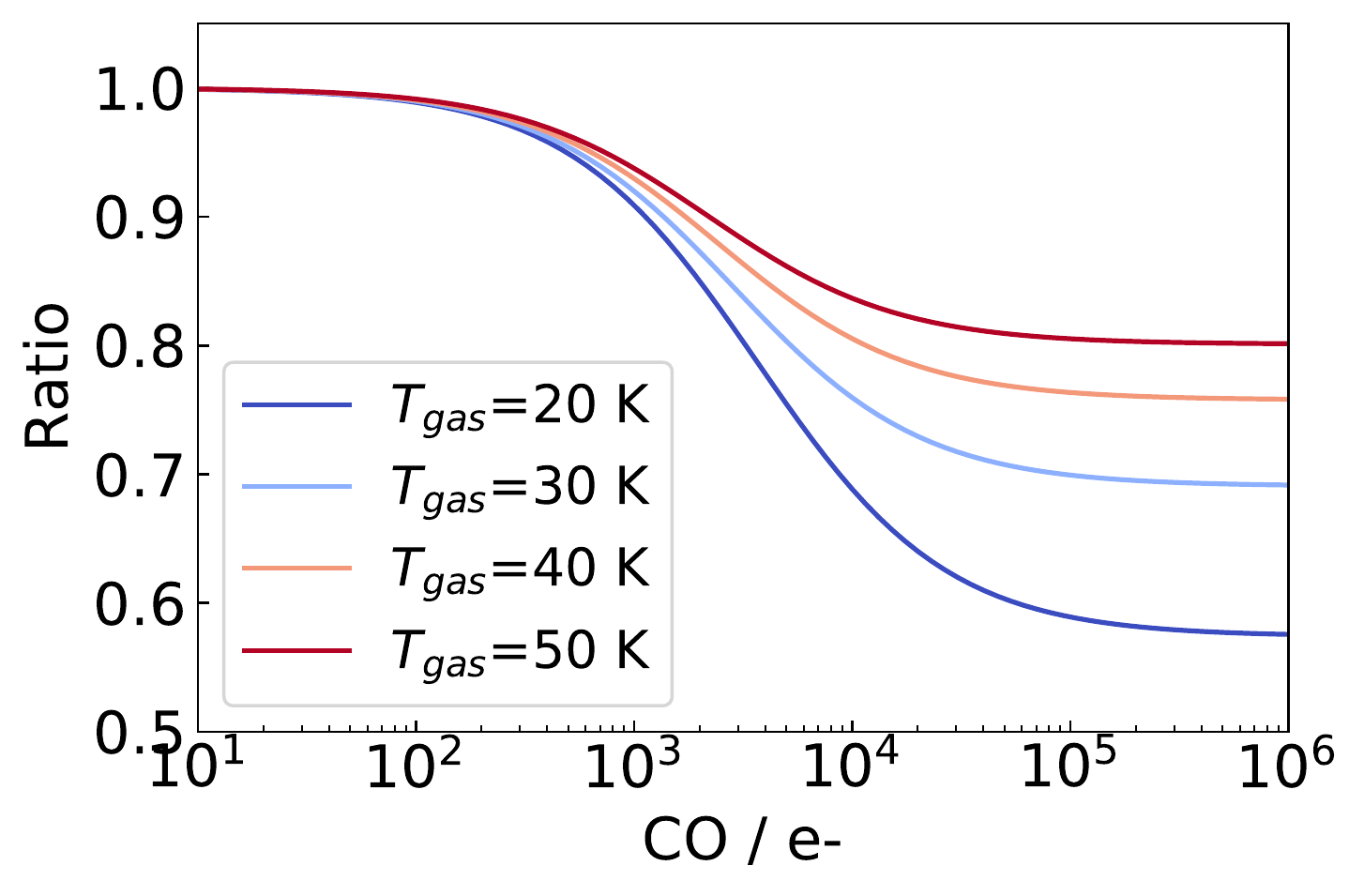}
    \caption{Ratio of \ce{^13CO}/\ce{C^18O} to \ce{H^13CO+}/\ce{HC^18O+} as a function of the abundance ratio between CO and electrons,
    varying the gas temperature. See the text for details.}
    \label{fig:analytical}
\end{figure}

Considering the above mentioned reactions and assuming the balance between the formation and destruction of the \ce{HCO+} isotopologues,
the relation between \ce{^13CO}/\ce{C^18O} and \ce{H^13CO+}/\ce{HC^18O+} is given by
\begin{align}
\frac{\ab{^{13}CO}}{\ab{C^{18}O}} =  \frac{k_{\rm ex1}\ab{CO} + k_{\rm rec}\ab{e^-}}{k_{\rm ex2}\ab{CO} + k_{\rm rec}\ab{e^-}} \times \frac{\ab{H^{13}CO^{+}}}{\ab{HC^{18}O^{+}}},  \label{eq:relation}
\end{align}
where $x(i)$ is the abundance of species $i$, $k_{\rm rec}$ is the rate coefficient of the electron recombination of the \ce{HCO+} isotopologues, and 
$k_{\rm ex1}$ and $k_{\rm ex2}$ are the rate coefficient in the reverse direction of Reaction \ref{react:13c} and Reaction \ref{react:18o}, respectively.
If both Reactions \ref{react:13c} and \ref{react:18o} are in chemical equilibrium, i.e., 
when the \ce{HCO+} destruction by the dissociative recombination with electrons is much slower than the isotope exchange reactions, Eq. \ref{eq:relation} is reduced to be  
$\ab{^{13}CO}/\ab{C^{18}O}$ = $\ab{H^{13}CO^{+}}/\ab{HC^{18}O^{+}} \times \exp(-11.1\,\, {\rm K}/T_{\rm gas})$.
Figure \ref{fig:analytical} visualizes the ratio of $\ab{^{13}CO}/\ab{C^{18}O}$ to $\ab{H^{13}CO^{+}}/\ab{HC^{18}O^{+}}$ as functions of $\ab{CO}/\ab{e^-}$.

It is difficult to constrain the $\ab{CO}/\ab{e^-}$ ratio from observations, because several species, such as \ce{H3+}, \ce{HCO+}, \ce{N2H+}, \ce{S+}, and \ce{C+}, would contribute to the total charge in the warm molecular layer \citep[e.g.,][]{aikawa15,teague15}. 
However, the $\ab{CO}/\ab{HCO^+}$ ratio would provide a rough upper limit of the $\ab{CO}/\ab{e^-}$ ratio in the warm molecular layer.
Assuming the \ce{HCO+}/\ce{H^13CO+} ratio of 69, the averaged \ce{HCO+} column density in the central 240  au $\times$ 110  au (4.1"$\times$1.8") regions is $\sim 4 \times 10^{13}$ cm$^{-2}$.
According to \citet{zhang19}, the CO column density at $r = 60-120$ au is $\sim$10$^{17}$--10$^{18}$ cm$^{-2}$.
Then the $\ab{CO}/\ab{HCO^+}$ ratio is roughly estimated to be $\sim3\times10^3$--$3\times10^4$.
The HCO$^{+}$ isotopologues emission would mainly originate from the regions at $T_{\rm gas} \gtrsim 20$ K in the TW Hya disk,
because HCO$^{+}$ is formed from CO and CO is frozen out onto dust grains at $\lesssim$20 K.
Taken together, we expect that the \ce{^{13}CO}/\ce{C^{18}O} ratio is 70--100 \% of the  \ce{H^{13}CO^{+}}/\ce{HC^{18}O^{+}} ratio.



\subsection{Model description}
To explore the relation between $\ab{^{13}CO}/\ab{C^{18}O}$ and $\ab{H^{13}CO^{+}}/\ab{HC^{18}O^{+}}$ in the TW Hya disk in more details,
we construct a physical and chemical model of the TW Hya disk with carbon and oxygen isotope fractionation chemistry.
The dust and gas density profiles in the disk are taken from the TW Hya disk model developed in \citet{cleeves15}.
The total gas mass is 0.04 $M_{\odot}$ and the dust-to-gas mass ratio is 0.01 as a whole.
Two populations of dust grains are considered: 
the small dust population with radii of 0.005 -- 1 $\mu$m and the large dust population with radii of 0.005 $\mu$m -- 1 mm.
The fraction of mass in the large dust population is set to be $f = 0.9$.
Both populations follow a power law size distribution with the index of -3.5.
The small dust population is assumed to be coupled with gas,
while the large dust population is concentrated near the midplane and its scale height is a factor of five smaller than that of gas \citep{cleeves15}.

The dust temperature in the disk is calculated using the radiative transfer code RADMC-3D \citep{dullemond12}.
The input UV and X-ray spectra of TW Hya are taken from \citet{dionatos19}, and 
they are added to a Black body component with the effective temperature of 4110 K \citep{andrews12}.
The local UV radiation and X-ray radiation fields inside the disk are also calculated with RADMC-3D, considering both absorption and scattering.
Our model does not consider the resonant scattering of Ly$\alpha$ photons, which dominates the stellar FUV flux \citep{herczeg02}, by atomic H \citep{bethell11b}.
This simplification would not affect our results significantly, as CO is not dissociated by Ly$\alpha$.  
The dust opacity for UV and longer wavelengths is calculated with \texttt{dsharp\_opac} package from \citet{birnstiel18}.
The gas and dust opacities for X-rays are calculated with the method proposed by \citet{bethell11} 
with the X-ray absorption and scattering cross sections taken form \texttt{xraylib} library \citep{brunetti04,schoonjans11}.

The gas temperature and chemical composition of the disk is calculated by 
integrating the energy equation and chemical rate equations simultaneously for 1 Myr,
which is shorter than the age of TW Hya \citep[$8\pm4$ Myr; ][]{donaldson16}, but long enough for the abundances of CO and \ce{HCO+} isotopologues to reach steady-state in our model.
The use of a shorter time than the age of TW Hya would be reasonable, 
as the current physical structure adopted represents the TW Hya disk today 
and not during the disk's entire 8 Myr evolution.
As heating and cooling processes, photoelectric heating, X-ray and cosmic-ray heating, heating by the \ce{H2} formation and photodissociation,
the energy exchange between gas and dust particles due to collisions, and 
line cooling by atoms and molecules (H, \ce{H2}, C, O, CO, OH, and \ce{H2O}) are considered \citep[e.g.,][]{tielens05}. 
The disk physical structures are shown in Appendix.

The chemical network used in this work is based on that in \citet{furuya14},
which includes gas-phase reactions, interaction between gas and (icy) grain surfaces, and grain surface reactions.
The network is extended to include mono-\ce{^13C} and mono-\ce{^18C} species (species with both \ce{^13C} and \ce{^18O}, e.g., \ce{^13C^18O}, are included), 
and carbon and oxygen isotope exchange reactions \citep{roueff15,mladenovic17,loison19,loison20}.
UV photodissociation/photoionization rates are calculated by convolving the local radiation field 
and the photodissociation/photoionization cross sections \citep{heays17}.
The self-shielding and mutual shielding factors for the photodissociation of CO isotopologues are taken from \citet{visser09}. 
The column densities for calculating the shielding factors are evaluated by the minimum of the inward/upward column.
This method is computationally cheap and often adopted in the literature \citep[e.g.,][]{miotello14}, 
but can underestimate the effect of self-shielding \citep[see the appendix of][]{lee21}.
X-ray chemistry is calculated in the same way as in \citet{furuya13}.
The binding energy of CO is set to be 855 K \citep{obegr05}.
The cosmic-ray ionization rate is set to be 10$^{-18}$ s$^{-1}$.

Table \ref{table:init_ab} lists initial molecular abundances for the disk model.
As shown by previous observations and the analysis based on thermo-chemical disk models, 
the CO abundance is low in the warm molecular layers ($>$20 K) and the C/O ratio is higher than unity in the TW Hya disk 
\citep[e.g.,][]{bergin16,kama16,zhang19,cleeves21,nomura21}.
In our model, the elemental carbon abundance is $1.4\times10^{-6}$ with the C/O ratio of 1.5.
As long as C/O $>$ 1, our model results are not sensitive to the choice of the C/O ratio, 
as the abundances of CO and \ce{HCO+} are regulated by the oxygen abundance rather than the carbon abundance.
Initially, the \ce{^12C}/\ce{^13C} and \ce{^16O}/\ce{^18O} ratios for all the species are set to be 69 and 557, respectively \citep{wilson99},
i.e., there is no fractionation at the beginning of the simulations. 


\begin{table}
\caption{Initial abundances with respect to hydrogen nuclei.}
\begin{center}
\begin{tabular}{cccc}
\hline\hline        
Species & Abundance & Species & Abundance  \\
\hline
\ce{H2} & 0.5 & N & 2.8(-5)  \\
He & 9.55(-2) & \ce{S+} & 8.0(-8)  \\
CO & 1.4(-6) & \ce{Si+} & 8.0(-9)  \\
C & 2.2(-6) & \ce{Fe+} & 3.0(-9) \\
\ce{H2O} & 1.0(-6) & \ce{Na+} & 2.0(-9) \\
\ce{N2} & 1.4(-5) & \ce{Mg+} & 7.0(-9) \\
\hline
\end{tabular}
\end{center}
\tablecomments{a(-b) means $a \times 10^{-b}$.}
\label{table:init_ab}
\end{table}

\subsection{Model results}
\begin{figure*}[ht!]
\plotone{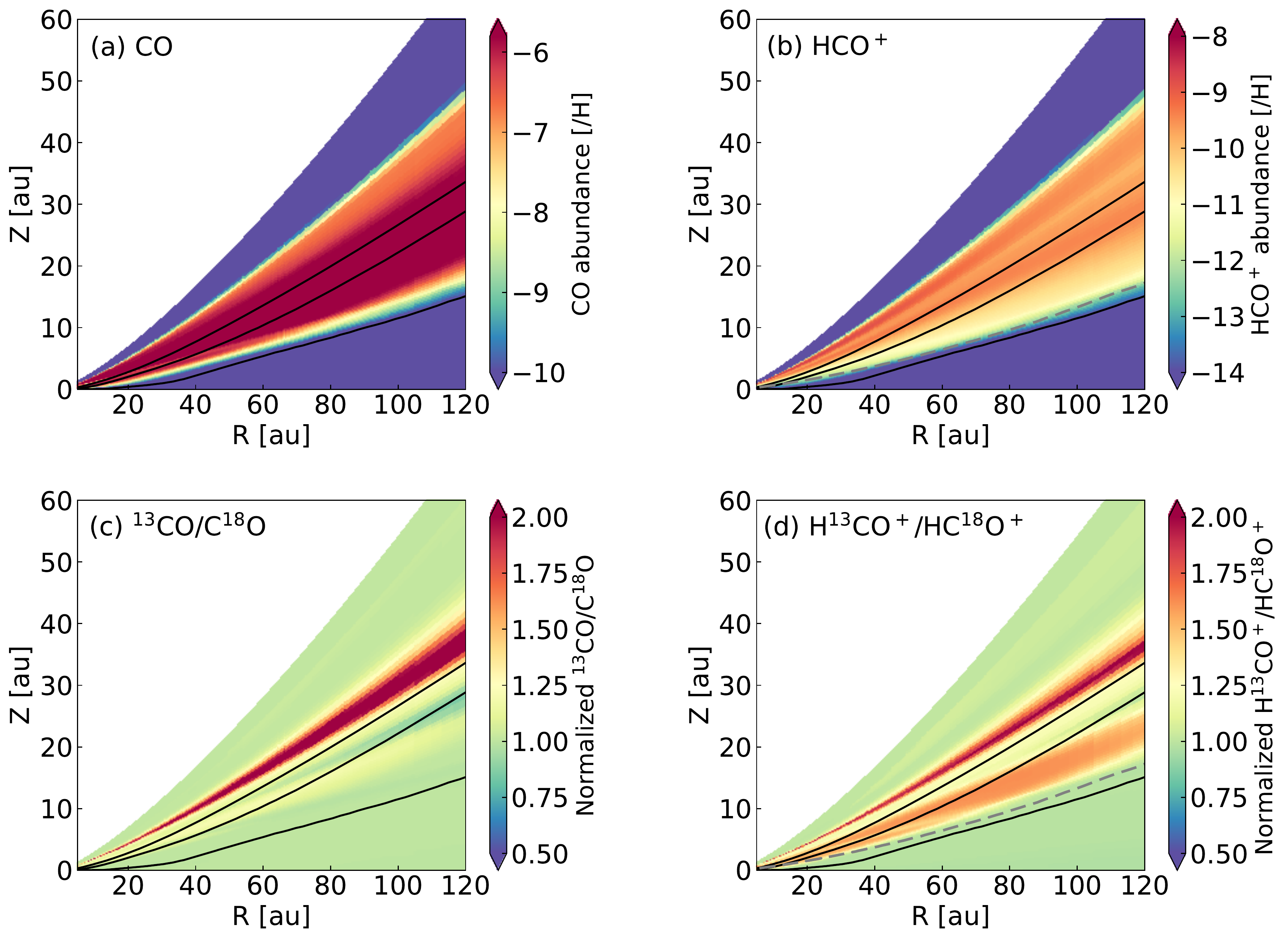}
\caption{Spatial distributions of fractional abundances of CO (panel a) and \ce{HCO+} (panel b) with respect to hydrogen nuclei, and 
the normalized abundances ratios of \ce{^13CO}/\ce{C^18O} (panel c) and \ce{H^13CO+}/\ce{HC^18O+} (panel d) with respect to the elemental abundance ratio of 8.1 in the disk model.
Black lines depict the positions where the gas temperature is equal to 20 K, 30 K, and 40 K.
Gray dashed lines in the panels b and d depict the positions where the X-ray ionization rate 
is equal to the cosmic-ray ionization rate (10$^{-18}$ s$^{-1}$).
}
\label{fig:model_ab}
\end{figure*}

The upper panels of Figure \ref{fig:model_ab} shows the spatial distributions of fractional abundances of CO and \ce{HCO+} with respect to hydrogen nuclei.
CO is abundant ($\sim$10$^{-6}$) in the warm molecular layers ($\gtrsim$20 K).
Above the warm molecular layers, CO is efficiently destroyed by UV photodissociation,
while below the warm molecular layers, CO is frozen out onto dust grains.
The spatial distribution of \ce{HCO+} basically follows that of CO, as it is formed by CO + H$_3^+$.
The dominant ionization sources in the regions where \ce{HCO+} is abundant are X-rays rather than cosmic-rays in our model, being consistent with the model in \citet{cleeves15}.

The lower panels of Figure \ref{fig:model_ab} shows the \ce{^13CO}/\ce{C^18O} abundance ratio and the \ce{H^13CO+}/\ce{HC^18O+} abundance ratio.
The \ce{^13CO}/\ce{C^18O} ratio does not show significant fractionation except for the thin layer ($z/R \sim 0.2-0.3$), 
where the ratio is higher than the elemental abundance ratio.
In these regions, \ce{C^18O} is not self-shielded, while \ce{^13CO} is, resulting in the overabundance of \ce{^13CO} with respect to \ce{C^18O}.
Indeed, the overabundance of \ce{^13CO} with respect to \ce{C^18O} has been observed in molecular clouds 
illuminated by UV photons from nearby stars \citep[e.g.,][]{shimajiri14}.
The \ce{H^13CO+}/\ce{HC^18O+} ratio is similar to the \ce{^13CO}/\ce{C^18O} ratio except for regions at $z/R \sim 0.1-0.2$,
where the abundance ratio between CO and electrons is $\sim$10$^4$-10$^5$, and Reactions \ref{react:13c} and \ref{react:18o} modify the \ce{H^13CO+}/\ce{HC^18O+} ratio (see Figure \ref{fig:analytical}).

\begin{figure*}[ht!]
\plotone{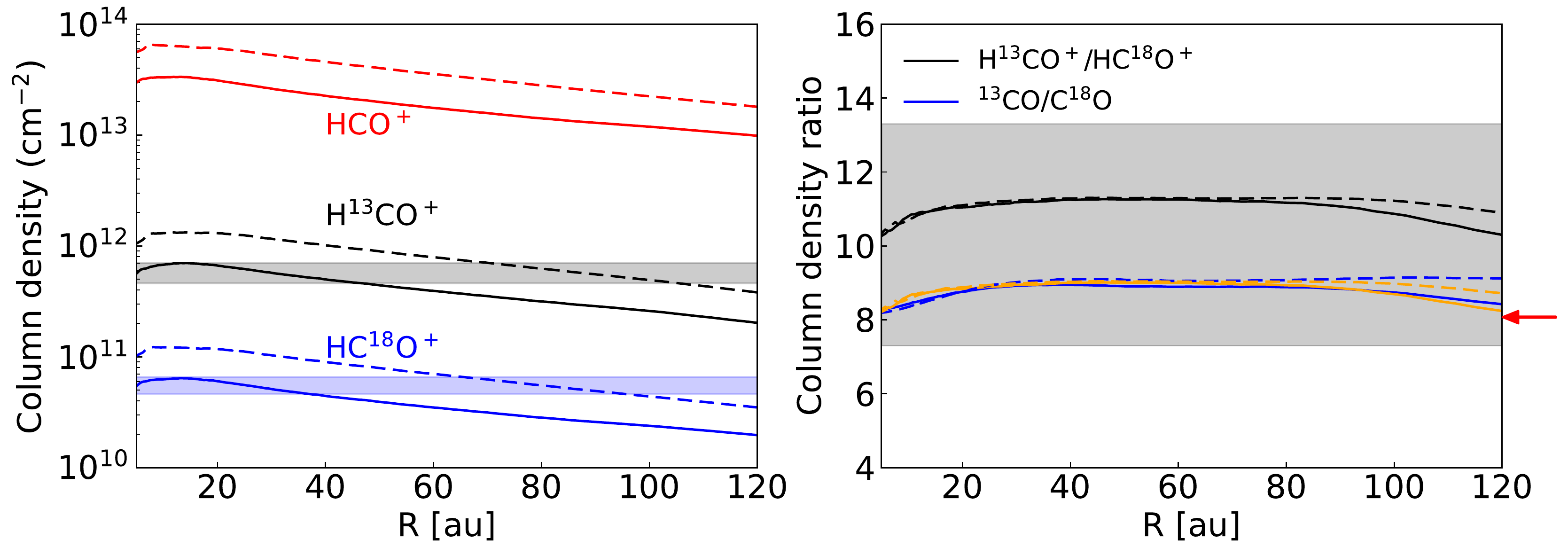}
\caption{Left) Radial profiles of the column densities of \ce{HCO+} isotopologues in our disk models. 
Right) Radial profiles of \ce{^13CO}/\ce{C^18O} column density ratio  (blue) and \ce{H^13CO+}/\ce{HC^18O+} column density ratio (black) in the disk models.
Orange line in the right panel represents the \ce{^13CO}/\ce{C^18O} column density ratio multiplied by a factor of 0.8.
Solid lines show the results of our fiducial model, while dashed lines show the results of the model with five times higher stellar X-ray flux.
Areas represent the observationally derived values in the TW Hya disk assuming $T_{ex} = 30$ K.
In the right panel, red arrow in the right margin shows the elemental [\ce{^13C}][\ce{^16O}]/([\ce{^12C}][\ce{^18O}]) ratio adopted in the disk model.
}
\label{fig:model_col}
\end{figure*}

The left panel of Figure \ref{fig:model_col} shows radial profiles of the column densities of \ce{HCO+} isotopologues.
The column densities of \ce{H^13CO+} and \ce{HC^18O+} in our model reasonably well reproduce those derived from the observations assuming $T_{ex} = 30$ K and the optically thin emissions.
We checked the optical depth of the \ce{H^13CO+}(4--3) line for a slab of \ce{H2} and 
\ce{H^13CO+} gas using the Radex code \citep{vandertak07}.
The peak \ce{H^13CO+} column density in our model is $7\times10^{11}$ cm$^{-2}$.
Assuming an \ce{H2} density of 10$^9$ cm$^{-3}$, the kinetic temperature of 30 K, linewidth of 0.13 km/s due to the thermal broadening, the \ce{H^13CO+} column density of $7\times10^{11}$ cm$^{-2}$ corresponds to the optical depth of $\sim$0.48 for the \ce{H^13CO+}(4--3) line.
Then as claimed by \citet{cleeves15}, the \ce{H^13CO+}(4--3) line is likely optically thin in the TW Hya disk.

The right panel of Figure \ref{fig:model_col} shows radial profiles of \ce{^13CO}/\ce{C^18O} column density ratio and \ce{H^13CO+}/\ce{HC^18O+} column density ratio.
The \ce{^13CO}/\ce{C^18O} column density ratio is very close to the elemental abundance ratio of 8.1,
because the isotope selective photodissociation of CO is significant only in limited regions.
This result is qualitatively consistent with the results of generic disk models in \citet{miotello14,miotello16}, 
who found that the impact of isotope selective photodissociation of CO is less significant in more massive disks with smaller CO abundance.
The \ce{H^13CO+}/\ce{HC^18O+} ratio is lower than \ce{^13CO}/\ce{C^18O} at all the radii in our disk model by a factor of $\sim$0.8.
Then in the rest of this paper, we assume the following relation from these models;
\ce{^13CO}/\ce{C^18O} = 0.8 $\times$ (\ce{H^13CO+}/\ce{HC^18O+}).


\section{Discussion \& Conclusion} \label{sec:discuss}
Given TW Hya's proximity, 
we can compare the isotopic ratio to the local ISM rato of [\ce{^13C}][\ce{^16O}]/([\ce{^12C}][\ce{^18O}]) of $8.1\pm0.8$ \citep{wilson99}.
Based on our measurement and model-derived correction factor, 
we estimate the \ce{^13CO}/\ce{C^18O} ratio in the TW Hya disk ratio to be $8.3 \pm 2.6$.
The \ce{^13CO}/\ce{C^18O} ratio is consistent with the elemental ratio in the local ISM.

\begin{figure}[ht!]
\plotone{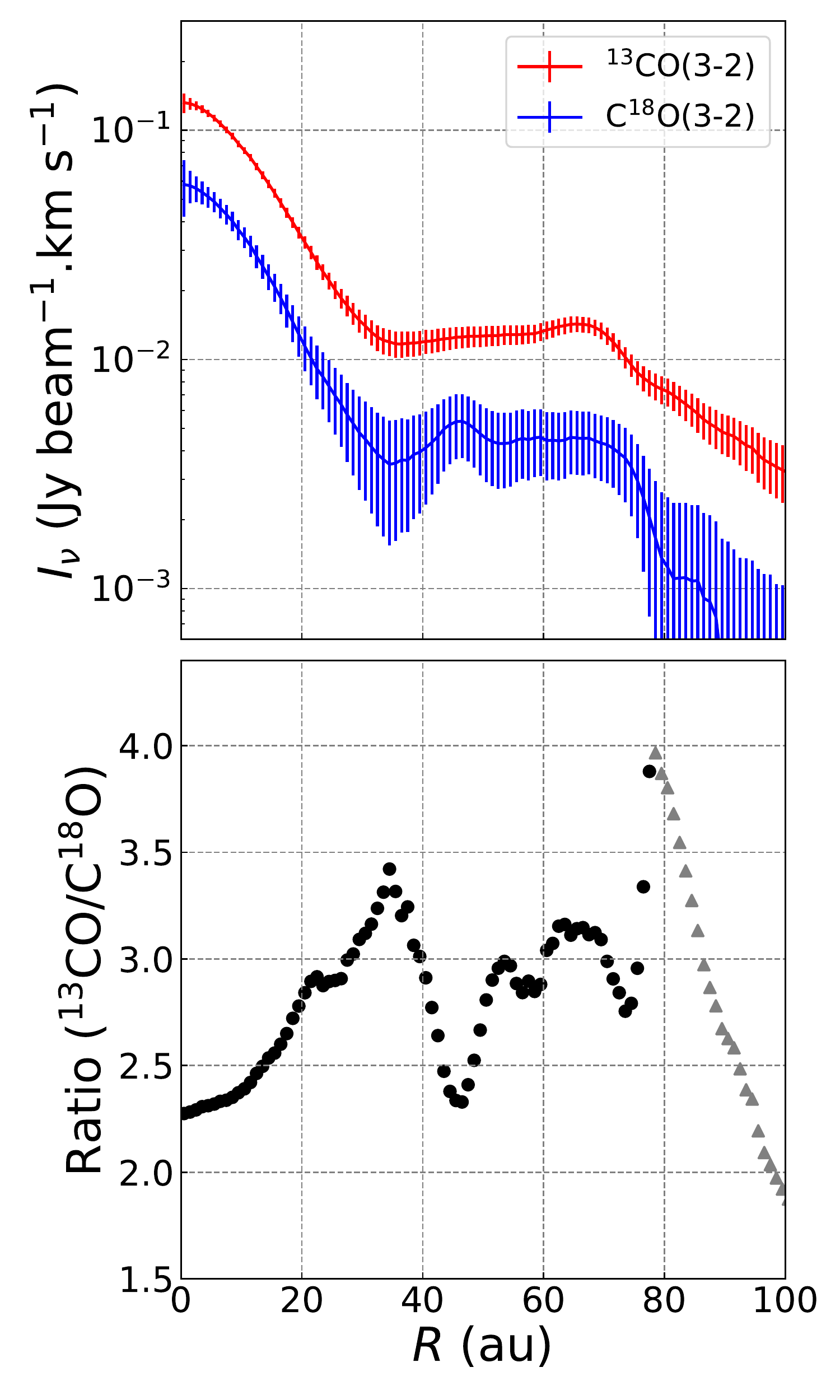}
\caption{
(Top) Radial profiles of the integrated intensity in \ce{^13CO}(3--2) and \ce{C^18O}(3--2) \citep{nomura21}.
(Bottom) Ratio of the \ce{^13CO}(3--2)/\ce{C^18O}(3--2) integrated intensities. 
The ratio is estimated for the \ce{C^18O} signals above 1.5$\sigma$ and shown in black points.
The gray triangles indicate the lower limit of the ratio calculated from the 1.5$\sigma$ value of the \ce{C^18O} emission.
}
\label{fig:ratio}
\end{figure}

It would be instructive to check how the \ce{^13CO}/\ce{C^18O} column density ratio 
estimated from \ce{H^13CO+}/\ce{HC^18O+} is compared 
with the \ce{^13CO}/\ce{C^18O} intensity ratio observed in the disk.
For this purpose, we use the ALMA data targeting \ce{^13CO}(3--2) and \ce{C^18O}(3--2) in the TW Hya disk 
with the spatial resolution of $\sim$9 au, presented in \citet{nomura21}.
The top and bottom panels of Figure \ref{fig:ratio} show the radial profiles of the integrated intensity of \ce{^13CO} and \ce{C^18O} and the intensity ratio of \ce{^13CO}/\ce{C^18O}, respectively. 
The \ce{^13CO}/\ce{C^18O} intensity ratio is as low as three at $R < 80$~au, indicating the \ce{^13CO} emission is 
likely optically thick as suggested by previous studies \citep{schwarz16}.
At optically thinner region beyond 80~au, the ratio may be larger than $\sim$4, 
but we cannot constrain the intensity ratio well due to the weak \ce{C^18O} emission.

Based on the \ce{C^18O}(3-2) and \ce{^13C^18O}(3-2) observations, 
\citet{zhang17} tentatively found that the \ce{CO}/\ce{^13CO} ratio in the TW Hya disk is $40^{+9}_{-6}$,
which is lower than the \ce{^12C}/\ce{^13C} ratio of 69 in the local ISM \citep{wilson99}.
The authors noted that higher sensitivity observations are needed to confirm this ratio, 
because the ratio was derived from the disk regions where the S/N ratio of \ce{^13C^18O}(3-2) was not high.
Our model predicts that bulk CO is slightly enriched in \ce{^13C} 
due to the isotope exchange reaction, \ce{^13C+} + \ce{CO} $\rightleftharpoons$ \ce{C+} + \ce{^13CO} + 35\,\,{\rm K} \citep[e.g.,][]{langer84}; 
the \ce{^12CO}/\ce{^13CO} column density ratio in our model is 62--69, depending on radius.
If we assume that \ce{^12CO}/\ce{^13CO} = 69,
the \ce{CO}/\ce{C^18O} ratio is estimated to be 390--750.
If we assume the \ce{^12CO}/\ce{^13CO} ratio derived by \citet{zhang17},
the estimated \ce{CO}/\ce{C^18O} ratio is lower than 530.

\epsscale{1.0}
\epsscale{1.0}

Although the \ce{CO}/\ce{C^18O} ratio in the proto-solar disk is unknown, there are some constrains on the H$_2^{16}$O/H$_2^{18}$O ratio.
Evidence for primordial heavy isotope enriched \ce{H2O} ice comes from Fe-O-S-bearing cosmic symplectites within Acfer 094 meteorite, 
which shows 25 \% enrichment of \ce{^18O} (and \ce{^17O}) compared to the Solar wind \citep{sakamoto07},
corresponding to H$_2^{16}$O/H$_2^{18}$O = 400.
Cometary water ice shows variations in H$_2^{16}$O/H$_2^{18}$O, ranging 400-500 \citep{bockelee-morvan15,altwegg19}.
Then, assuming the \ce{H2O}/CO abundance ratio of unity at the moment when oxygen isotope fractionation was implemented, 
CO should be depleted in \ce{^18O} (and \ce{^17O}) by up to 25 \% compared to the Solar wind,
corresponding to \ce{CO}/\ce{C^18O} = 660.
The disk chemical model with turbulent mixing \citep{lyons05} predicted that CO could be depleted in \ce{^18O} by up to $\sim$60 \% compared to the Solar wind, corresponding to \ce{CO}/\ce{C^18O} = 850.
Note that the [\ce{^16O}]/[\ce{^18O}] ratio in the Solar wind and that in the local ISM is similar (530 versus $557 \pm 30$).


Taken together, at this moment, there is no clear evidence of \ce{^18O} depletion in the observable CO gas in the central 100 au regions of the TW Hya disk.
However, there are uncertainties, including the large error in the \ce{H^13CO+}/\ce{HC^18O+} ratio (and thus \ce{^13CO}/\ce{C^18O}), 
and we cannot draw any robust conclusions yet.
If confirmed, this implies that the material evolution in the TW Hya disk is different from that experienced in the proto-solar disk for some reason.
One possibility might be the different birth environment (the TW Hya disk is in quiescent regions, while the sun was likely born in a cluster region),
i.e., external UV radiation filed might be the crucial parameter for the oxygen isotope fractionation scenario.
Future higher sensitivity observations of \ce{HCO+} isotopologues are necessary for better constraining the oxygen isotope ratio of CO gas in the TW Hya disk.
The confirmation of the low \ce{^12CO}/\ce{^13CO} ratio suggested by \citet{zhang17} is also crucial for constraining the oxygen isotope ratio.



Finally, we note that the possible effect of stellar X-ray variability on the above discussion.
\citet{cleeves17} found that \ce{H^13CO+}(3-2) line intensity from the disk around a Tauri star, IM Lup, measured in May 2015 
is brighter by a factor of $\sim$2 compared to that in January 2015, due to X-ray flare events.
As the \ce{HC^18O+} data and \ce{H^13CO+} data used in this work were taken at a different epoch,
we can not rule out the possibility that the \ce{H^13CO+}/\ce{HC^18O+} ratio, 
and thus the \ce{^13CO}/\ce{C^18O} ratio derived in this work, is affected by stellar X-ray variability.
In order to check the effect of X-ray variability quickly, we reran our disk model with five times higher stellar X-ray flux.
We confirmed that the column densities of the \ce{HCO+} isotopologues are increased by a factor of $\sim \sqrt{5}$, because the main ionization source is X-rays in the regions where \ce{HCO+} is abundant in our model, and because the abundance of major molecular ions, such as \ce{HCO+}, is proportional to $\sqrt{\xi}$, where $\xi$ is the ionization rate \citep[e.g.,][]{aikawa15}.
On the other hand, the column density ratios of \ce{H^13CO+}/\ce{HC^18O+} and \ce{^13CO+}/\ce{C^18O} 
do not change significantly (the dashed lines in Figure \ref{fig:model_col}).
Then, simultaneous observations of \ce{HCO+} isotopologues would be able to eliminate the possible effect of X-ray variability on the measurements of the isotopic ratios of \ce{HCO+} and thus those of CO.

\acknowledgments
This work makes use of the following public ALMA archive data: ADS/JAO.ALMA\#2016.1.00311.S.
ALMA is a partnership of ESO (representing its member states), NSF (USA) and NINS (Japan), together with NRC (Canada), MOST and ASIAA (Taiwan), and KASI (Republic of Korea), in cooperation with the Republic of Chile. 
A part of the data analysis was carried out on the common-use data analysis computer system at the Astronomy Data Center of NAOJ.
Numerical computations were in part carried out on PC cluster at Center for Computational Astrophysics, NAOJ.
K.F. is supported by JSPS KAKENHI Grant numbers 20H05847, 21H04487, and 21K13967.
T.T. is supported by JSPS KAKENHI Grant number 20K04017.

\begin{appendix}
\section{Disk Physical Structure}
\begin{figure}[ht!]
\plotone{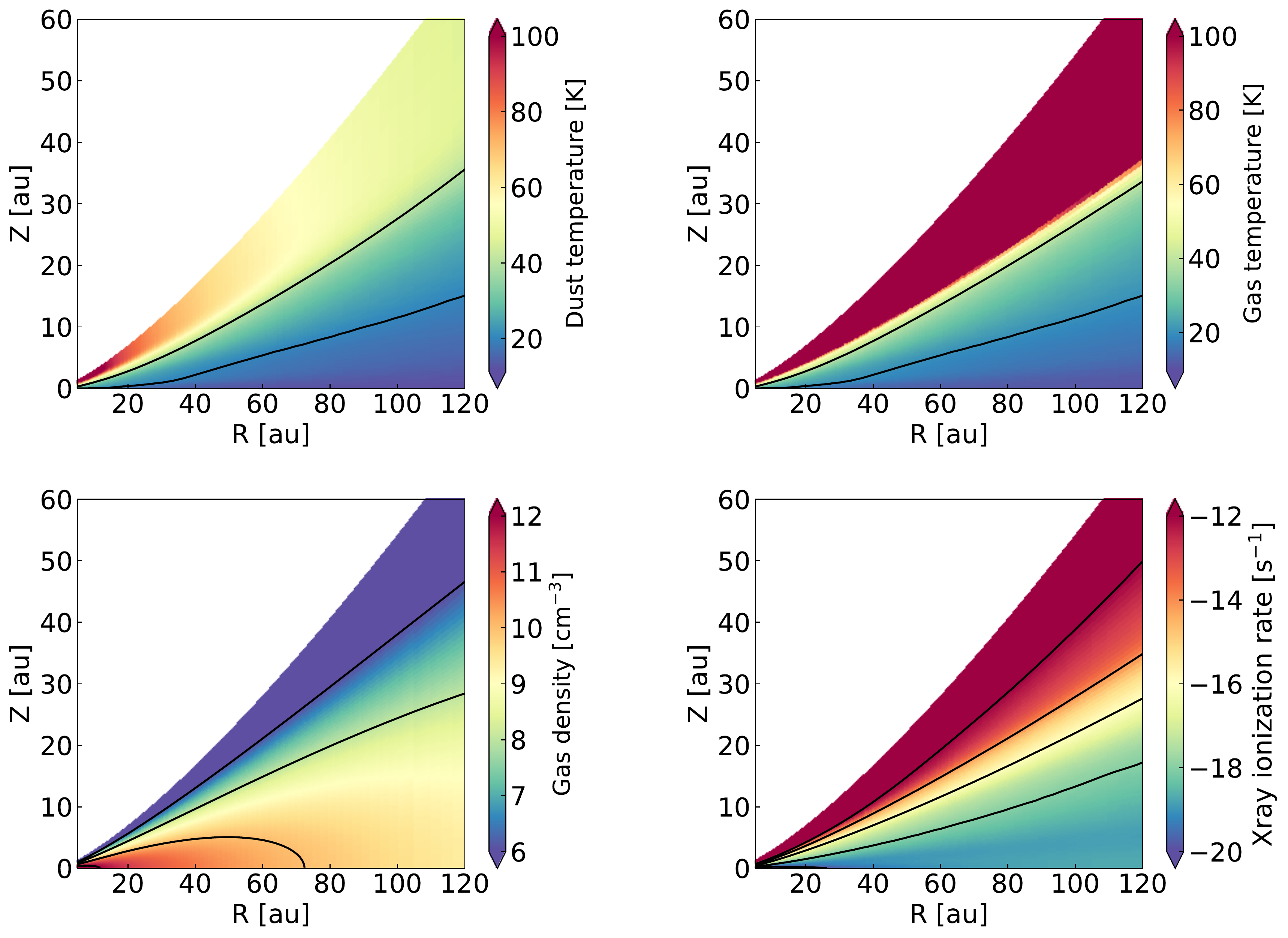}
\caption{Spatial distributions of dust temperature, gas temperature, gas density, and X-ray ionization rate in our disk model.
}
\label{fig:append}
\end{figure}

Figure \ref{fig:append} shows the disk physical structures in our model.

\end{appendix}

\bibliography{ms_hc18o+}{}
\bibliographystyle{aasjournal}





\end{document}